\documentclass[11pt]{amsart}
\usepackage[utf8]{inputenc}
\usepackage{physics}
\usepackage{mathtools,amsmath,amsfonts,amssymb,amsthm}
\usepackage{hyperref}
\usepackage{xcolor}
\usepackage{comment}
\usepackage{stmaryrd}
\usepackage{enumerate}
\usepackage{soul} 

\newcommand\C{\mathbb{C}}
\newcommand\N{\mathbb{N}}

\newcommand\R{\mathbb{R}}
\newcommand\HH{\mathcal{H}}
\newcommand\F{\mathbb{F}}
\newcommand\Fq{\mathbb{F}_q}

\newcommand\RM{\mathrm{RM}}
\newcommand\ie{\textit{i.e.,~}}
\newcommand\eg{\textit{e.g.,~}}
\newcommand{\wt}{\mathrm{wt}}
\newcommand{\short}{\mathrm{short}}
\newcommand{\punct}{\mathrm{punct}}

\theoremstyle{plain}
\newtheorem{theorem}{Theorem}[section]
\newtheorem{corollary}[theorem]{Corollary}
\newtheorem{proposition}[theorem]{Proposition}
\newtheorem{lemma}[theorem]{Lemma}
\theoremstyle{definition}
\newtheorem{definition}[theorem]{Definition}
\newtheorem{remark}[theorem]{Remark}
\newtheorem{example}[theorem]{Example}

\usepackage[margin=3cm]{geometry}

\title{Structure of CSS and CSS-T Quantum Codes}

\author{Elena Berardini}
\address{CNRS; IMB, Université de Bordeaux, 351 cours de la Libération, 33405 Talence, France}
\email{elena.berardini@math.u-bordeaux.fr}
\author{Alessio Caminata}
\address{Dipartimento di Matematica, Dipartimento di Eccellenza 2023-2027, Universit\`a di Genova\\ via Dodecaneso 35, 16146, Genova, Italy}
\email{alessio.caminata@unige.it}
\author{Alberto Ravagnani}
\address{Eindhoven University of Technology, The Netherlands}
\email{a.ravagnani@tue.nl}

\keywords{Quantum error-correcting code,
 CSS code, CSS-T code, code parameters}
\date{}

\usepackage{setspace}
\begin{document}

\begin{abstract}
We investigate CSS and CSS-T quantum error-correcting codes from the point of view of their 
existence, rarity, and performance.
We give a lower bound on the number of pairs of linear codes that give rise to a CSS code with good correction capability, showing that such pairs are easy to produce with a randomized construction. We then prove that CSS-T codes exhibit the opposite behaviour, showing also that, under very natural assumptions, their rate and relative distance cannot be simultaneously large.
This partially answers an open question on the feasible parameters of CSS-T codes.
We conclude with a simple construction of CSS-T codes from Hermitian curves.
The paper also offers a concise introduction to CSS and CSS-T codes from the point of view of classical coding theory.
\end{abstract}

\maketitle

\section*{Introduction}

Quantum computers are 
inherently susceptible to errors and disturbances.
Quantum error correction is a crucial aspect of quantum computing. It is about protecting quantum information from errors due to decoherence and other forms of noise; see~\cite{nat,unruh1995maintaining} among many others. Quantum error correction is currently a wide-open challenge.

In 1996, Calderbank and Shor \cite{CS96}, and independently Steane \cite{S96}, 
proposed a class of
quantum codes, mostly known under the name of \textit{CSS codes}, obtained by 
combining two classical linear error-correcting codes. Since then, several articles have studied their construction
and obtained quantum codes using known families of linear codes, such as \textit{Reed--Solomon}  and \textit{BCH codes} \cite{GGB99,LG14}, \textit{Reed--Muller codes} \cite{SK05,S99}, and \textit{Algebraic Geometry codes} \cite{HMMM20,JX11,KM08}.

Quantum CSS codes are most often constructed by taking the two classical codes needed for the construction to be a
\textit{self-orthogonal} code and its dual.
On the other hand, this approach is technically not required by the construction and, as we will argue in this paper, imposes strong constraints.

Very recently, Rengaswamy \emph{et al.}~\cite{RCNP20,RCNP20b} introduced a class of CSS codes, called \textit{CSS-T codes}, specifically designed for universal fault-tolerant quantum computation. The properties of CSS-T codes are to date largely unexplored. An open question is about the existence of families of CSS-T codes whose rate and relative distance are both non-vanishing for large block length. This paper provides some partial answers.

In the rest of the introduction, we briefly summarize the contributions made by this paper and point the reader to the relevant sections.
In Section~\ref{sec1}, we provide the necessary background material, which also gives us a chance to concisely present quantum error-correcting codes from the point of view of
classical coding theory.
In Section~\ref{sec2}, we study the parameters of CSS codes and give a lower bound on the number of code pairs that give rise to a CSS code with sufficiently large correction capability. In particular, we prove that over a large field, a randomized construction will produce a CSS code with good parameters (Corollary~\ref{cor:density}). This builds on preliminary results on this topic derived in~\cite{lenka}. 

Section~\ref{sec3} is devoted to CSS-T codes. We provide a convenient characterization using results on finite bilinear spaces and then apply this to derive bounds on their parameters (Theorem~\ref{prop:CSST}). In particular, we show that CSS-T codes cannot have large rate and relative distance simultaneously (Theorem~\ref{th:csstbounds}). This makes some progress on an open question about the existence of asymptotically good families of CSS-T codes. Moreover, it proves that the construction proposed in~\cite{RMCSST} is optimal. Finally, in Section~\ref{sec4}, we show how some simple examples of CSS-T codes can be constructed as evaluation codes from the Hermitian curve (Theorem~\ref{th:hermitianCSST}). This construction is optimal with respect to the previously mentioned bounds on rate and relative distance.

\subsection*{Acknowledgements}
The authors would like to thank Felice Manganiello for 
introducing them to CSS-T codes, Paolo Solinas for useful discussions on quantum channels, and Markus Grassl for pointing out some mistakes in an earlier version of this paper. We also thank the anonymous referees for the valuable comments and suggestions.

\section{Quantum Error-Correcting Codes}\label{s:QECC}\label{sec1}
In this section, we give a self-contained, very short introduction to the theory of quantum error-correcting codes. This also gives us the chance to establish the notation for the rest of the paper.
We refer the reader to \cite{BCH20,Grassl21,NRS06,NC02,R19,Scherer19} for more details.
\subsection{Qudits}\label{ss:qudits}
Let $q$ be the power of a prime $p$ and let $\Fq$ denote the finite field with $q$ elements. Let $\C$ denote the complex field, $\dagger$ the conjugate transpose of a matrix, and $I$ the $q\times q$ identity matrix over $\C$. Matrices and vectors will be denoted by uppercase and lowercase letters, respectively. The all-zero and all-one vectors are denoted by $\boldsymbol{0}$ and $\boldsymbol{1}$, respectively.

We denote by $b_0,\dots, b_{q-1}$ the elements of $\Fq$ and let $\ket{b_0},\ket{b_1},\dots \ket{b_{q-1}}$ represent unit vectors in $\C^q$ as follows:
\begin{align*}
    \ket{b_0}= \begin{pmatrix}
    1 \\
    0 \\
    \vdots \\
    0 \\
    0
    \end{pmatrix}, \qquad 
    \ket{b_1}= \begin{pmatrix}
    0 \\
    1 \\
    0 \\
    \vdots \\
    0
    \end{pmatrix}, \qquad \dots \qquad
    \ket{b_{q-1}}= \begin{pmatrix}
    0 \\
    0 \\
    \vdots \\
    0 \\
    1 
    \end{pmatrix}.
\end{align*}
The symbol $\ket{\cdot}$ is read
``ket''. Note that $\ket{b_0},\ket{b_1},\dots \ket{b_{q-1}}$ are orthonormal unitary vectors. We can represent each vector $\alpha\in\C^q$ with respect to the previous basis
\begin{equation}
\label{quantumstate2}
    \ket{\alpha} \coloneqq \sum_{i=0}^{q-1}\alpha_i\ket{b_i},
\end{equation}
where $\alpha_i \in \C$.  We also introduce the following notation, called the “bra":
\[\bra{\alpha}\coloneqq \sum_{i=0}^{q-1}\overline{\alpha_i}\bra{b_i},\]
where $\bra{b_i}$ is the conjugate transpose of $\ket{b_i}$, \ie $\bra{b_i}=\ket{b_i}^\dagger$. Finally, we let
$\mathcal{H}_q$ be the Hilbert space $\C^q$
\begin{equation}\label{Hilbert_q}
\mathcal{H}_q\coloneqq \left\{\sum_{i=0}^{q-1} \alpha_i \ket{b_i}\mid \alpha_i \in \C\right\}
\end{equation}
endowed with the scalar product 
\[\bra{\alpha}\ket{\beta}\coloneqq\sum_{i=0}^{q-1} \overline{\alpha_i}\beta_i,\]
and the corresponding norm 
$\lVert\ket{\alpha}\rVert\coloneqq\sqrt{\bra{\alpha}\ket{\alpha}}.$

A \emph{qudit}  is an element of $\HH_q$ of norm one. 
The coefficients $\alpha_i$'s in Equation \eqref{Hilbert_q} are called the \emph{probability amplitudes} of $\ket{\alpha}$. From the point of view of quantum physics, this means that the particle represented by the qudit $\ket{\alpha}$ is in the state $\ket{b_i}$ with probability $|\alpha_i|^2$.
For $q=2$, an element of $\HH_2$ of norm one is called a \emph{qubit}.

We define the \emph{tensor product} of   $\ket{\alpha}$ and $\ket{\beta}$ as the usual tensor product of the associated vectors in $\C^q$. Note that the tensor product is associative but not commutative. 
For $\ket{x_1},\dots,\ket{x_n}\in\HH_q$, we let
$\ket{x_1\dots x_n}$ denote the tensor product $\ket{x_1}\otimes\dots\otimes\ket{x_n}$.
We then set
\[\HH_q^{\otimes n}\coloneqq \mathcal{H}_q\otimes \dots \otimes \mathcal{H}_q,
\]
and sometimes omit $q$ when it does not need to be specified (we write simply $\HH^n$).
The space $\HH^n$ is a $q^n$-dimensional Hilbert space.
 The elements of the form $\ket{b}$ with $b \in \Fq^n$  form a basis of $\HH^n$, which is called the \emph{computational basis}. 

\begin{example}
$\HH_3^2$ is spanned by $\{\ket{00},\ket{01},\ket{02},\ket{10},\ket{11},\ket{12},\ket{20},\ket{21},\ket{22}\}$. The kets in the basis we gave correspond to the nine vectors of the standard basis of $\C^9$, \eg
\[\ket{02}\coloneqq\ket{0}\otimes\ket{2}=\begin{pmatrix}1\\ 0\\0\end{pmatrix}\otimes\begin{pmatrix}0\\ 0\\1\end{pmatrix}=\begin{pmatrix}0\\0\\1\\ 0\\0\\0\\0\\0\\0\end{pmatrix}.\]
\end{example}

A system of $n$-qudits is represented as the tensor product of $n$ qudits.  Thus a $n$-qudit is a vector of the form $\smash{\ket{\alpha}=\sum_{b\in \Fq^n} \alpha_b \ket{b}\in\HH^ n}$ with $\smash{\sqrt{\sum_{b\in \Fq^n} |\alpha_b|^2} =1}$. It is important to notice that not every $n$-qudit can be written as a tensor product of $n$ qudits. Indeed, this is only possible when the qudits are independent, or \emph{uncorrelated}. A correlation between two qudits is called an \emph{entaglement}.

\begin{example}
The $2$--qubits
\[\frac{1}{\sqrt{2}}(\ket{01}+\ket{10}) \text{ and } \frac{1}{\sqrt{2}}(\ket{00}+\ket{11})\]
cannot be written as tensor products of two qubits. They are \emph{entangled} qubits.
\end{example}

 \subsection{Quantum Errors and Quantum Codes}
The goal of this subsection is to introduce the possible errors that can occur in the quantum realm. For simplicity of exposition, we will start with the case of qubits.
Quantum errors are better explained by using the language of density matrices.

\begin{definition}[{\cite[Postulate 5]{Scherer19}}]
A complex $2\times 2$ matrix $\rho$ is a called a \emph{density matrix} if  the following properties hold:
\begin{itemize}
\item $\rho$ is a Hermitian (or self-adjoint) operator, \textit{i.e.}, $\rho=\rho^\dagger$,
    \item $\rho$ is positive semi--definite, \textit{i.e.},~$\bra{\alpha}\rho\ket{\alpha}\geq 0$ for every $\alpha\in\mathbb{C}^2$,
    \item $\Tr(\rho)=1$.
\end{itemize}
\end{definition}

Given a qubit $\ket{\alpha}=\alpha_0 \ket{0}+\alpha_1\ket{1}$, we can  represent it as a density matrix $\rho_{\alpha}$ in the following way:
\[\rho_{\alpha}=\ket{\alpha}\bra{\alpha}=\begin{pmatrix}
|\alpha_0|^2 & \alpha_0\overline{\alpha_1}\\
\overline{\alpha_0}\alpha_1 & |\alpha_1|^2
\end{pmatrix}.\]

Notice that not all density matrices $\rho$ are associated to a qubit $\ket{\alpha}$, that is $\rho=\rho_{\alpha}$. When this happens, we say that the density matrix is in \emph{pure state}, otherwise, we say that $\rho$ is in \emph{mixed state}. Alternatively, a density matrix in mixed state can be thought of as representing a set of qubits $\ket{\alpha_i}$ with associated probabilities $p_i$, such that $\sum_i p_i=1$.
The nature of a state of a density matrix $\rho$ is completely captured by $\rho^2$, in the following way \cite[Proposition~2.25]{Scherer19}.
\begin{proposition}\label{prop:purestaterho}
The density matrix $\rho$ represents a pure state if, and only if, $\rho=\rho^2$.
\end{proposition}

Now, suppose that we send a density matrix $\rho_{\alpha}$ representing a qubit $\ket{\alpha}$ through a channel. Some errors may occur, but the received output is still a density matrix $\rho'$. This condition corresponds to the physical fact that under any circumstances we should be able to measure the output $\rho'$ and obtain the value $0$ or $1$ with some probabilities $p_1$, $p_2$ such that $p_1+p_2=1$. However, the matrix $\rho'$ needs not to be in pure state anymore, but it may be in mixed state. 
Quantum errors are modeled as linear transformations \cite{Grassl21}. In the literature, one often refers to them also as \emph{gates} or \emph{operators}.
\begin{definition}[\protect{\cite[Definition 2.1]{Grassl21}}]
A \emph{quantum channel} $\mathcal{E}$ is a linear transformation acting on density matrices:
    \[\mathcal{E}: \rho\mapsto \mathcal{E}(\rho)=\sum_i E_i \rho E_i^\dagger.\]
The $E_i$'s are called \emph{(quantum)  error operators} or \emph{Krauss operators}.
\end{definition}
We have infinitely many possible quantum errors. 
However, thanks to the linearity of the operators, it is sufficient to be able to correct the errors corresponding to the Pauli basis of the space of $2 \times 2$  matrices (see the discussion in~\cite[\S 2.2]{R19} and \cite[\S 2.2]{Grassl21}): 
$$I, \qquad 
\sigma_x=\begin{pmatrix} 0 & 1\\1&0
\end{pmatrix}, \qquad 
\sigma_y=i\begin{pmatrix} 0 & -1\\1&0
\end{pmatrix}, \qquad 
\sigma_z=\begin{pmatrix} 1 & 0\\0&-1
\end{pmatrix}.$$
Besides the identity matrix, the Pauli matrices are sometimes written with the following alternative notation, which we will often use in the sequel:
\begin{equation}\label{eq:Paulimatrices}
X=\sigma_x, \qquad Z=\sigma_z, \qquad  Y=XZ=-i\sigma_y.
\end{equation}
Precisely, they are referred to as:
\begin{itemize}
    \item the $X$--error (bit flip):  $\alpha_0\ket{0}+\alpha_1\ket{1} \longrightarrow \alpha_0\ket{1}+\alpha_1\ket{0}$,
    \item the $Z$--error (phase flip): $\alpha_0\ket{0}+\alpha_1\ket{1}
    \longrightarrow \alpha_0\ket{0}-\alpha_1\ket{1}$,
    \item the $Y$--error (phase and bit flip): $\alpha_0\ket{0}+\alpha_1\ket{1}
    \longrightarrow
    \alpha_0\ket{1}-\alpha_1\ket{0}$.
\end{itemize}
Notice that the Pauli matrices do not commute.
\begin{remark}
Using the Pauli matrices as a basis of the space of $2\times 2$ matrices, we can write any density matrix $\rho$ as
\[\rho=\frac{1}{2}(I+x \sigma_x+y \sigma_y + z \sigma_z),\]
for some $x,y,z\in\R$.
By Proposition~\ref{prop:purestaterho}, $\rho$ is in pure state, \ie $\rho=\rho_\alpha$ for some qubit $\ket{\alpha}$, if and only if  $\rho^2=\rho$. In particular, this implies $\Tr(\rho^2)=1$ which is equivalent to saying that $|(x,y,z)|=1$.
Therefore we can represent the qubit $\ket{\alpha}$ as an element of the real sphere $\{(x,y,z)\in\mathbb{R}^3\mid \ x^2+y^2+z^2=1\}$.
\end{remark}

\begin{example}
 We fix the notation convention that quantum gates are applied from right to left. Let $\ket{\alpha}=\alpha_0\ket{0}+\alpha_1\ket{1}$.
 Then 
\[XZ\ket{\alpha}=\alpha_0\ket{1}-\alpha_1\ket{0}\neq -\alpha_0\ket{1}+\alpha_1\ket{0}=ZX\ket{\alpha}.\]
\end{example}

\begin{example}\label{ex:hadamard}
Another important qubit unitary operation is the \emph{Hadamard gate} $H$, obtained by dividing the classical Hadamard matrix by its determinant. This results in a unitary matrix that acts on the computational basis as follows:
\[H\ket{0}=\frac{1}{\sqrt{2}}(\ket{0}+\ket{1}),\quad H\ket{1}=\frac{1}{\sqrt{2}}(\ket{0}-\ket{1}).\]
Its matrix form is
\[H=\frac{1}{\sqrt{2}}\begin{pmatrix}1 & 1\\1&-1
\end{pmatrix}=\frac{1}{\sqrt2}\left(\sigma_x+\sigma_z\right).\]
The Hadamard gate is the inverse of itself: $HH^{-1}=I$. In fact, it can be interpreted as a linear change of basis on the space $\HH$: from the computational basis $\left\{\ket{0}, \ket{1}\right\}$ to the basis  $\left\{H\ket{0}, H\ket{1}\right\}$.
\end{example}

For quantum errors over $\mathcal{H}_q$  with $q=p^e>2$ and $p$ a prime, the situation is very similar to the case $q=2$. We can associate to a qudit $\ket{\alpha}$ a $q\times q$ density matrix $\rho_{\alpha}=\ket{\alpha}\bra{\alpha}$ as before, and 
we can consider the errors corresponding to the higher-dimensional analogue of the Pauli matrices. 
The equivalent to the bit-flip error is the \emph{dit-flip}. For each element $a \in \F_q$, this can be represented by a $q \times q$ matrix $X(a)$ sending $\ket{\alpha}$ to $\ket{\alpha + a}$. For the equivalent of the phase-flip error, the so-called \emph{phase-shift}, we define $Z(b)$ to be a $q \times q$ diagonal matrix such that its $i$-th entry on the diagonal is $\zeta^{\Tr_{\F_q / \F_p} (ib)}$ for each $b \in \F_q$. Here $\zeta = e^{2\pi i/p}\in\mathbb{C}$ is a primitive $p$-th root of unity, and $\Tr_{\F_q / \F_p}$ is the usual \emph{absolute trace} defined as follows:
\begin{equation}\label{eq:absolutetrace}
\Tr_{\F_q / \F_p} (\alpha)= \sum_{i=0}^{e-1} \alpha^{p^i} .
\end{equation}
For any $a,b\in\Fq$, the unitary matrices $X(a)$ and $Z(b)$ are called the \emph{Pauli matrices}. When the dependency on $a,b$ is not relevant, we shall denote them simply by $X$ and $Z$ respectively. Notice that over $\F_2$ they coincide with the Pauli matrices given in Equation \eqref{eq:Paulimatrices}. We refer the reader to \cite[\S 5]{BCH20} for further details.

\begin{example}
For $q=3$, the dit-flip errors are
$$
X(0) = \begin{bmatrix}
1 & 0 & 0 \\
0 & 1 & 0 \\
0 & 0 & 1
\end{bmatrix},
\qquad
X(1) = \begin{bmatrix}
0 & 1 & 0 \\
0 & 0 & 1 \\
1 & 0 & 0 
\end{bmatrix},
\qquad
X(2) = \begin{bmatrix}
0 & 0 & 1 \\
1 & 0 & 0 \\
0 & 1 & 0 
\end{bmatrix},
$$ 
while the phase-shift errors are
$$
Z(0) = \begin{bmatrix}
1 & 0 & 0 \\
0 & 1 & 0 \\
0 & 0 & 1
\end{bmatrix},
\qquad
Z(1) = \begin{bmatrix}
1 & 0 & 0 \\
0 & \zeta & 0 \\
0 & 0 & \zeta^2
\end{bmatrix},
\qquad
Z(2) = \begin{bmatrix}
1 & 0 & 0 \\
0 & \zeta^2 & 0 \\
0 & 0 & \zeta
\end{bmatrix}.
$$
\end{example}

\begin{example}\label{ex:qhadamard}
    Let $\omega$ be a primitive $q$-th root of unity. The $q$-dimensional equivalent of the Hadamard gate introduced in Example~\ref{ex:hadamard} is the $q$-dimensional discrete Fourier transform matrix, which reads
    $$H^{(q)}=\frac{1}{\sqrt{q}}\begin{bmatrix}
1 & 1 & 1 & \dots & 1 \\
1 & \omega & \omega^2 &  \dots & \omega^{q-1} \\
1 & \omega^2 & \omega^4 &  \dots & \omega^{2(q-1)} \\
\vdots &\vdots & \vdots & \ddots &\vdots \\
1 & \omega^{q-1} &\omega^{2(q-1)}&\dots & \omega^{(q-1)(q-1)}\\
\end{bmatrix}.$$
Notice that for $q$ a power of $2$, we obtain the unitary $q$-dimensional Hadamard matrix. 
The matrix $H^{(q)}$ can be seen as a linear change of basis on the space $\HH_q$.
\end{example}

As in the classical setting, the vast majority of the literature focuses on the case of \emph{memoryless quantum channel}, meaning that the noise acts independently in each channel use. Therefore, we can represent the $q^n\times q^n$ matrices acting on $n$-qudits as $n$-tensors of $q\times q$ matrices. For $e,f\in\Fq^n$, with $e=(e_1,\dots,e_n)$ and $f=(f_1,\dots,f_n)$, we use the following notation:
\begin{equation}\label{eq:errors}
X^eZ^f=\bigotimes_{i=1}^n X(e_i)Z(f_i).
\end{equation}

The qudits in a $n$-qudit are implicitly labeled in increasing order from $1$ to $n$. Consequently, one can put indices on the unitary matrices to indicate which qudit they act on or assume by default that the $i$-th matrix acts on the $i$-th qudit. 
\begin{example}
The $3$-qubit $\ket{010}$ corresponds to $\ket{0}_1\otimes\ket{1}_2\otimes\ket{0}_3$. Writing
\[X_1Z_2I_3\ket{010}=XZI\ket{010}=-\ket{110}\]
means that the Pauli matrix $X$ acts on the first qubit, the Pauli matrix $Z$ acts on the second, while the third qubit is unchanged. The identity matrix at the end can be omitted, \ie we can simply write $X_1Z_2\ket{010}$ or $XZ\ket{010}$.
\end{example}

\begin{definition}[\protect{\cite[Definition~7.12]{Scherer19}}]
An \emph{encoding} is an injective and norm-preserving linear operator $\Phi:\HH_q^{\otimes k}\rightarrow\HH_q^{\otimes n}$. The space $\HH_q^{\otimes n}$ is called the \emph{quantum encoding space}, and the image of the encoding $\mathcal{Q}=\mathrm{Im}\Phi$ is called a \emph{quantum error-correcting code} (\textit{QECC}) of dimension $k$ and length $n$. 
For brevity, we say that $\mathcal{Q}$ is a \emph{$\llbracket n,k\rrbracket_q$-quantum code}. The elements of $\mathcal{Q}$ are called codewords.
\end{definition}

\begin{example}[Steane code]
The Steane code was introduced in \cite{S96}. It is a $\llbracket 7,1\rrbracket_2$-quantum code. The encoding is given by the linear map $\Phi:\HH_2\rightarrow\HH_2^{\otimes 7}$ defined by
\[
\begin{split}
\ket{0}\mapsto\frac{1}{\sqrt8}\big(&\ket{0000000}+\ket{1010101}+\ket{0110011}+\ket{1100110}\\
+&\ket{0001111}+\ket{1011010}+\ket{0111100}+\ket{1101001}\big), \\
\ket{1}\mapsto\frac{1}{\sqrt8}\big(&\ket{1111111}+\ket{0101010}+\ket{1001100}+\ket{0011001}\\
+&\ket{1110000}+\ket{0100101}+\ket{1000011}+\ket{0010110}\big).
\end{split}
\]
\end{example}

In contrast to the classical setting of linear codes with the Hamming distance, we do not have a general decoding procedure associated with a quantum error-correcting code. In fact, a quantum error-correcting code would be better specified by giving an encoding and a decoding algorithm \cite[Definition~7.12]{Scherer19}. In this paper, we are interested only in CSS codes and their variants, which have an associated known decoding algorithm \cite{CS96}. In particular, we don't treat decoding in this paper.

\section{CSS Codes: Parameters and Randomized Constructions}\label{s:CSS}\label{sec2}
CSS codes, named after Calderbank, Shor, and Steane, are a class of error-correcting quantum codes derived from a pair of classical linear codes. Their construction was introduced by Calderbank and Shor~\cite{CS96}, and independently by Steane~\cite{S96}, in 1996. The two constructions are different but turn out to be equivalent. The original construction of CSS codes uses binary classical linear codes. However, it can be easily extended to codes over any finite field~$\Fq$. In the sequel, we focus on this more
general setting. CSS codes have been extensively studied since their introduction, and consequently, there is a rich literature about them. Besides the many constructions cited in the introduction to this paper, one can find an improved version of CSS codes in \cite{S99CSS}, a geometric interpretation of CSS codes (and more generally of stabilizer codes) in \cite{BCH20}, and some new constructions of CSS quantum codes in \cite{G23}.
\smallskip

The main goal of this section is to study the density property of CSS codes (Subsection~\ref{ss:densityCSS}). 
We will show that under mild hypotheses any pair of linear codes give a quantum CSS code with large minimum distance (Theorem \ref{th:CSSbound} and Corollary \ref{cor:density}). In particular, good CSS codes can be obtained with a randomized construction. Before doing so, we will 
introduce CSS codes and their parameters in Subsection \ref{ss:CSScodes}. For 
completeness, we will also briefly recall both the construction by Calderbank and Shor, and the one by Steane, and show their equivalence in the non-binary setting. We were not able to find a complete proof of this fact in the literature. This section builds on~\cite{lenka}, which contains the mentioned density result.

Before diving into the CSS code construction, we establish the notation
for classical linear codes. We work over a finite field  $\Fq$, where $q$ is a power of prime $p$. A (\emph{linear}) \emph{code}
$C$ is a linear subspace of $\F_q^n$, where $n$ is a positive integer. The vectors $c\in C$ are called \emph{codewords}. The \emph{dimension} of a code $C$ is its dimension as a linear subspace over $\F_q$. If a linear code $C_2$ is a subspace of a linear code $C_1$, we 
say that $C_2$ is a \emph{subcode of} $C_1$.

The \emph{support} of a vector $c \in \F_q^n$  is the set 
$
\sigma(c):=\{i \in \{1,\dots, n\} \, | \, c_i \neq 0\}$.
Its (\emph{Hamming}) \emph{weight} is 
$\wt(c) := |\sigma(c)|$, \ie the cardinality of the support of $c$.
The \emph{minimum distance} of a code~$C$
is defined as $$ d(C)=\min\{\wt(c) \mid c \in C, \, c \neq \boldsymbol{0}\},$$ where the zero code $\{\boldsymbol{0}\}$ has minimum distance $n+1$ by definition.
The \emph{dual} of $C$ 
is the code
\[C^\perp\coloneqq\{ x\in\Fq^n \mid \langle c,x\rangle=0 \mbox{ for all } c \in C\},\]
where $\langle c,x \rangle$ denotes the inner product of the vectors $c$ and $x$.
If $C$ has dimension $k$, then $C^\perp$ has dimension $n-k$. A code $C$ is said to be \emph{self-orthogonal} if $C\subseteq C^\perp$, and \emph{self-dual} if $C=C^\perp$.

\subsection{Parameters of CSS Codes}\label{ss:CSScodes}
 Fix linear codes $C_2 \subseteq C_1\subseteq \F_q^n$.
 Let $\zeta=e^{2\pi i/p}\in\mathbb{C}$ be a primitive $p$-th root of unity and let $\Tr_{\Fq / \F_p}=\Tr$ denote the absolute trace from $\Fq$ to its prime subfield $\F_p$ defined in Equation \eqref{eq:absolutetrace}.  
For a vector $w \in \F_q^n$ we define a quantum state as follows
\begin{equation}\label{CSqudit}
\ket{c_w} \coloneqq \frac{1}{\sqrt{|C_1|}}\sum_{c\in C_1}\zeta^{\Tr(\langle c,w \rangle)}\ket{c}.
\end{equation}
Note that for any $w \in \F_q^n$,  $\ket{c_w}$ has norm one and thus defines a qudit.
\begin{definition}
 The (Calderbank and Shor)  CSS quantum code associated with $C_1$ and $C_2$ is 
\[Q^{\textnormal{CS}}(C_1,C_2) = \{ \ket{c_w} \,|\, w \in C_2^\perp \}.\]
\end{definition}
In the previous definition, we used the superscript $\textnormal{CS}$ to distinguish this construction from the one of Steane~\cite{S96}, which we are going to present now.

Let $w\in C_1$. We define a qudit as 
\begin{equation}\label{eq:CSSqudit}
\ket{w + C_2} := \frac{1}{\sqrt{|C_2|}} \sum_{c\in C_2} \ket{w+c},
\end{equation}
where the $+$ in the ket denotes the usual coordinate-wise addition in $\F_q^n$. 

\begin{definition}
 The (Steane)  CSS quantum code associated with $C_1$ and $C_2$ is 
$$
Q^{\textnormal{S}}(C_1,C_2)=\{\ket{w+C_2} \,|\, w \in C_1 \}.
$$
\end{definition}

\begin{remark}
    It is not difficult to prove that $w-w' \in C_2 $ if and only if $\ket{w+C_2}=\ket{w'+C_2}$, and that $w-w'\in C_2$ if and only if the two cosets $w+C_2$ and $w'+C_2$ are equal. Hence, the natural index set for the qudits of the form $\ket{w+C_2}$ is the set of cosets $C_1 / C_2 $. Similarly, the natural index set for the qudits $\ket{c_w}$ is the set of cosets $C_2^\perp / C_1^\perp$. 
\end{remark}

As announced at the beginning of this section, the two definitions of CSS codes are equivalent, \ie they differ by a locally unitary operator. Indeed, the two constructions give a different basis for the same quantum code. This was pointed out by Calderbank and Shor in the binary case \cite{CS96}. 
Since we could not find a suitable reference for the general case, we present a proof for the analogous statement when $q>2$.
\begin{proposition}\label{prop:equivalenceCSS}
Let $C_2\subseteq C_1\subseteq\Fq^n$ be linear codes, and let $C_1^\perp$ and $C_2^\perp$ denote their respective dual codes. Then, the codes $Q^{\textnormal{CS}}(C_1,C_2)$ and $Q^{\textnormal{S}}(C_2^\perp,C_1^\perp)$ are equivalent.
\end{proposition}

\begin{proof}
Let $w\in C_2^\perp$. With the same notation as before, consider an element of $Q^{\textnormal{CS}}(C_1,C_2)$
\[
\ket{c_w} = \frac{1}{\sqrt{|C_1|}}\sum_{c\in C_1}\zeta^{\Tr(\langle c,w \rangle)}\ket{c}.
\]
The action of the unitary $q$-dimensional discrete Fourier transform matrix $H^{(q)}$ (see Example~\ref{ex:hadamard}) on an element $\ket{b}$ of the standard basis of $\HH_q$ is 
\[H^{(q)}\ket{b}\coloneqq \frac{1}{\sqrt{q}}\sum_{j=0}^{q-1}\zeta^{\Tr(j b)}\ket{j}.\]
Applying this to $\ket{c_w}$ gives
\[
    \ket{c'_w}\coloneqq H^{(q)}\ket{c_w}=\frac{1}{\sqrt{|C_1||\Fq^n|}}\sum_{z\in\Fq^n}\sum_{c\in C_1}\zeta^{\Tr(\langle c, (w+z)\rangle)}\ket{z}.
\]
Putting $z'=z+w$ in the previous equation gives
\begin{equation}\label{cw_change}
    \ket{c_w'}=\frac{1}{\sqrt{|C_1||\Fq^n|}}\sum_{z'\in\Fq^n}\sum_{c\in C_1}\zeta^{\Tr(\langle c, z'\rangle)}\ket{z'-w}.
\end{equation}
Recall that
\[
   \sum_{c\in C_1}\zeta^{\Tr(\langle c,z'\rangle)}= \begin{cases}
    |C_1|\text{ if } z'\in C_1^\perp,\\
    0 \text{ otherwise.}
    \end{cases}
\]
Therefore Equation \eqref{cw_change} can be rewritten as
    \begin{equation}\label{cw_final}
       \ket{c'_w}=\frac{|C_1|}{\sqrt{|C_1||\Fq^n|}}\sum_{z'\in C_1^\perp}\ket{z'-w} =\frac{1}{\sqrt{|C_1^\perp|}}\sum_{z'\in C_1^\perp}\ket{-w+z'},
\end{equation}
which is a codeword of $Q^{\textnormal{S}}(C_2^\perp,C_1^\perp)$, as defined in Equation \eqref{eq:CSSqudit}. 
Since $Q^{\textnormal{CS}}(C_1,C_2)$ and $Q^{\textnormal{S}}(C_2^\perp,C_1^\perp)$ have the same dimension, this concludes the proof.
\end{proof}

From now on we will not distinguish between the two (equivalent) definitions of CSS codes.
Given two linear codes $C_2 \subseteq C_1 \subseteq \mathbb{F}_q^n$ of dimension $k_1$ and $k_2$ respectively, we let $Q(C_1,C_2)=
Q^{\textnormal{CS}}(C_1,C_2)$ be the associated CSS code, which has parameters $\llbracket n, k_1-k_2\rrbracket_q$.

In order to discuss the quantum error correction capability of CSS codes, we first need to define the weight of an error as given in Equation \eqref{eq:errors} (see~\cite[Definition F.~63]{Scherer19}).
\begin{definition}
    For $e,f\in\Fq^n$, let $X^eZ^f=\bigotimes_{i=1}^{n} X(e_i)Z(f_i)$ be an error acting on a $n$-qudit. The weight of $X^eZ^f$ is defined as the number of tensor factors different from the identity, that is 
    \[\wt(X^eZ^f)\coloneqq \#\{i\in \{1,\dots,n\}\,|\, X(e_i)Z(f_i)\neq I\}=|\sigma(e)\cup\sigma(f)|.\]
\end{definition}
Note that the number of $X$ errors (respectively, $Z$ errors) in $X^eZ^f$ equals $\wt(e)$ (respectively, $\wt(f)$).

The following result
about the quantum error correction capability of $Q(C_1,C_2)$ holds.

\begin{theorem}[\cite{CS96}]\label{th:css}
Let $C_2 \subseteq C_1 \subseteq \mathbb{F}_q^n$  be two linear codes. Let $d_1$ and $d_2^\perp$ denote the minimum distance of $C_1$ and $C_2^\perp$, respectively. Then $Q(C_1,C_2)$ can correct any quantum error pattern $X^eZ^f$ for $\wt(e)\leq\lfloor\frac{d_1-1}{2}\rfloor$ and $\wt(f)\leq\lfloor\frac{d_2^\perp-1}{2}\rfloor$. 
\end{theorem}

Given the fact that quantum errors can be modeled as a pattern of errors of type $X$ and $Z$, the previous result shows that the minimum distances  $d_1$ and $d_2^\perp$ are relevant parameters of  $Q(C_1,C_2)$. In particular, if $d=\min\{d_1, d_2^\perp\}$, then 
$Q(C_1,C_2)$ can correct any error pattern of weight at most $\left\lfloor\frac{d-1}{2}\right\rfloor$. In this case, we say that $Q(C_1,C_2)$ is $\left\lfloor\frac{d-1}{2}\right\rfloor$-error correcting.

\begin{remark}
  The parameters $d_1$ and $d_2^\perp$ play a symmetric role in the error correction capability of $Q(C_1,C_2)$. Indeed,
  we can correct  $\lfloor(d_1-1)/2\rfloor$ errors of type $X$ and  $\lfloor(d_2^\perp-1)/2\rfloor$ errors of type $Z$. On the other hand, the CSS code $Q(C_2^\perp,C_1^\perp)$ is equivalent to $Q(C_1,C_2)$ by Proposition~\ref{prop:equivalenceCSS} and  can correct  $\lfloor(d_2^\perp-1)/2\rfloor$ errors of type $X$ and $\lfloor(d_1-1)/2\rfloor$ errors of type~$Z$.
\end{remark}

\subsection{Density Properties of CSS Codes}\label{ss:densityCSS}
By Theorem \ref{th:css}, for any pair of codes $C_2\subseteq C_1$ we can construct a quantum code $Q(C_1,C_2)$ which is (at least) $\left\lfloor\frac{d-1}{2}\right\rfloor$-error correcting, where $d=\min\{d_1,d_2^\perp\}$ and $d_1$ and $d_2^\perp$ denote the minimum distance of $C_1$ and $C_2^\perp$, respectively. In particular, if we choose $C_1=C^\perp$ such that $C\subseteq C^\perp$, and we take $C_2=C$, we obtain a CSS code $Q(C^\perp,C)$ which is $\left\lfloor\frac{d-1}{2}\right\rfloor$-error correcting, with $d=d_1=d_2^\perp$ being the minimum distance of $C^\perp$. For this reason, self-orthogonal codes have been extensively studied in relation to quantum codes. However, a CSS code which is $\left\lfloor\frac{d-1}{2}\right\rfloor$-error correcting for some fixed integer~$d$, can be constructed from any pair
of codes $C_2\subseteq C_1$ such that $d_1\geq d$ and $d_2^\perp\geq d$. 

The goal of this subsection is to estimate the number of pairs of classical linear codes $C_2\subseteq C_1$ such that $Q(C_1,C_2)$ is a quantum $t$-error-correcting code, with $t=\lfloor \frac{d-1}{2}\rfloor$ for some fixed $d$. We will show that, under some mild conditions on $d$, and for $q$ large enough, the set of such pairs is dense. In particular, 
focusing on self-orthogonal linear codes is extremely restrictive.

We briefly 
recall some preliminary notations and results.
Given a vector $x\in \F_q^n$,  the \emph{ball of radius $r\in \N$ of center $x$} is the set 
$
B(x,r):=\{ y \in \F_q^n \, | \, d(x,y)\leq r \}\subseteq \F_q^n
$.
We let
\begin{equation}\label{eq:boldb}
    \boldsymbol{b}(r)= |B(x,r)|= \sum_{i=0}^{r}(q-1)^i{n \choose i}.
\end{equation}
For positive integers $n\geq k$, their $q$-binomial coefficient is defined as 
$$
\binom{n}{k}_q = \frac{(1-q^n)(1-q^{n-1})\dots(1-q^{n-k+1})}{(1-q^k)(1-q^{k-1})\dots (1-q)},
$$
which is
the number of vector subspaces of dimension $k$ of a vector space of dimension $n$ over $\F_q$. We will repeatedly use the asymptotic estimate for the
$q$-binomial coefficient as $q$ grows, that is,
\[\binom{n}{k}_q\sim q^{k(n-k)}\quad\text{ as }q\rightarrow\infty.\]

\begin{proposition}[\protect{\cite[ Theorem~5.1]{BR20}}]\label{p:subset}\label{th:count} 
Let $C\subseteq \F_q^n$ be a linear code of dimension $k<n$ and let $h$ be an integer such that $k\leq h \leq n$. For an integer $2\leq d\leq d(C)$, set  \[\mathcal{F}(C)_{h,<d}:=\{D \subseteq \F_q^n \mid C \subseteq D, \, \dim(D)=h\text{ and }d(D)<d \}.\] Then
\[
|\mathcal{F}(C)_{h,<d}| \leq \binom{n-h}{h-k}_q \; \frac{q^{h}-q^{k}}{(q-1)(q^n-q^{k})} (\boldsymbol{b}(d-1)-1).
\]
\end{proposition}

In the following proposition, we show that two different pairs of linear codes give two different CSS codes.
Thus, in order to count the number of CSS codes $Q(C_1,C_2)$ with prescribed parameters it is enough to count pairs of codes $(C_1,C_2)$.

\begin{proposition}\label{pp:equivalence}  Let $C_2 \subseteq C_1\subseteq \F_q^n$ and $C_2' \subseteq C_1'\subseteq \F_q^n$ be linear codes. Let $Q(C_1,C_2)$ and $Q(C_1',C_2')$ denote the two associated CSS codes. Then $Q(C_1,C_2)=Q(C_1',C_2')$ if and only if $C_1=C_1'$ and $C_2=C_2'$.
\end{proposition}
\begin{proof}
The backward direction is obvious, so we focus on proving the forward direction. Assume that for codes $C_2\subseteq C_1 \subseteq \F_q^n$ and $C_2'\subseteq C_1' \subseteq \F_q^n$ we have $Q(C_1,C_2)=Q(C_1',C_2')$. Thus for every $\ket{c_w}\in Q(C_1,C_2)$ there is a unique quantum codeword $\ket{c_{w'}}\in Q(C_1',C_2')$ such that  $\ket{c_w}=\ket{c_{w'}}$, that is
\begin{equation}
\label{equationequality}
    \frac{1}{\sqrt{|C_2|}} \sum_{c\in C_2} \ket{w+c}=\frac{1}{\sqrt{|C_2'|}} \sum_{c'\in C_2'} \ket{w'+c'}.
\end{equation}
We have $\ket{w+c_1}=\ket{w+c_2}$ if and only if $c_1=c_2$, thus the $|C_2|$ kets in both sum are all different and correspond to a standard basis vector, hence the sums in Equation~\eqref{equationequality} contain $|C_2|$ and $|C_2'|$ distinct orthogonal vectors, respectively. Hence every $\ket{w+c}$ is equal to exactly one $\ket{w'+c'}$, implying that every $w+c$ is equal to exactly one $w'+c'$. Moreover, the sums contain the same number of elements, hence~$|C_2|=|C_2'|$.

We will first show that $C_1=C_1'$. Assume by contradiction that $C_1\neq C_1'$. Without loss of generality, we can assume that there exists an element $\bar{w}\in C_1\backslash C_1'$. By hypothesis there exists $w'\in C_1'$ such that $\ket{c_{\bar w}}=\ket{c_{w'}}$, that is for any $c\in C_2$ there exists $c'\in C_2'$ such that $\bar{w}+c=w'+c'$. Then, take $c=\boldsymbol{0}\in C_2\subseteq C_1$, so for some $c'\in C_2'$ we must have $\bar{w}=\bar{w}+\boldsymbol{0}=w'+c'\in C_1'$, a contradiction. Hence we must have $C_1=C_1'$.

Now, let us show that $C_2=C_2'$ holds too. We will prove this again by contradiction. Take $\ket{c_w}\in Q(C_1,C_2)$ and $\ket{c_{w'}}\in Q(C_1,C_2')$  such that $\ket{c_w}=\ket{c_{w'}}$. Taking $c=\boldsymbol{0}\in C_2$ gives $\ket{w+\boldsymbol{0}}=\ket{w'+c'}$ for some $c'\in C_2'$, which then implies $w=w'+c'$ and hence $w-w'\in C_2'$. Since $\ket{c_w}=\ket{c_{w'}}$, for all $c \in C_2$ there exists $c'\in C_2'$ such that $\ket{w+c}=\ket{w'+c'}$, which entails $c'-c=w-w'\in C_2'$. Therefore, we also have $c=c'-(c'-c) \in C_2'$ and thus $C_2\subseteq C_2'$, which together with $|C_2| = |C_2'|$ implies $C_2=C_2'$.
\end{proof}

We are now ready to establish the main result of this subsection. Let  $C_2 \subseteq C_1 \subseteq \mathbb{F}_q^n$ be linear codes. Let $d_1$ and $d_2^\perp$ denote the minimum distance of $C_1$ and $C_2^\perp$, respectively. If $d_1\geq \alpha$ and $d_2^\perp\geq \beta$, for some integers $\alpha$ and $\beta$, then by Theorem~\ref{th:css} we know that the quantum code $Q(C_1,C_2)$ is $\lfloor \frac{d-1}{2} \rfloor$-error correcting with $d\geq\min\{\alpha,\beta\}$. 
Our goal is to estimate the cardinality of the set 
\begin{equation}\label{setS'}
\mathcal{S}_{k_1,k_2}^{\alpha,\beta}\coloneqq\{(C_1,C_2)\in\mathcal{S}\mid d_1\geq \alpha, d_2^\perp\geq \beta\},
\end{equation}
as a subset of 
\begin{equation}\label{setS}
\mathcal{S}_{k_1,k_2}\coloneqq\{(C_1,C_2)\mid C_2\subseteq C_1, \, \dim C_1=k_1, \, \dim C_2=k_2\}.
\end{equation}
The second set has cardinality $|\mathcal{S}_{k_1,k_2}|=\binom{n}{k_1}_q\binom{k_1}{k_2}_q.$
In order to estimate the size of $\mathcal{S}_{k_1,k_2}^{\alpha,\beta}$ we introduce the following two auxiliary sets:
\[
\begin{split}
\mathcal{E}(\Fq^n)_{k_1,\geq \alpha}&\coloneqq\{C_1\subseteq \Fq^n \mid \dim(C_1)=k_1, \, d(C_1)\geq \alpha\},\\
\mathcal{G}(C_1)_{k_2,\geq \beta}&\coloneqq\{C_2\subseteq \Fq^n \mid C_2\subseteq C_1, \, \dim(C_2)=k_2, \, d(C_2^\perp)\geq \beta\}.
\end{split}
\]

\begin{lemma}\label{countingC1}
    Let $k_1\leq n$ and $\alpha$ be integers. We have
\[|\mathcal{E}(\Fq^n)_{k_1,\geq \alpha}|\geq\binom{n}{k_1}_q\left[1-\frac{q^{k_1}-1}{(q-1)(q^n-1)}(\boldsymbol{b}(\alpha-1)-1)\right].\]
\end{lemma}
\begin{proof}
    Note that $\mathcal{E}(\Fq^n)_{k_1,\geq \alpha}$ is the complement of 
    \[\mathcal{F}(\Fq^n)_{k_1,<\alpha}=\{C_1\subseteq \Fq^n\mid \dim(C_1)=k_1, d(C_1)< \alpha\}\]
in the set $\{C_1 \subseteq \F_q^n \, | \,  \dim(C_1)=k_1 \}$ of dimension $\binom{n}{k_1}_q$.
Applying Proposition \ref{th:count} to the set $\mathcal{F}(\Fq^n)_{k_1,<\alpha}$ gives
\begin{equation}\label{eq:Gbis}
    |\mathcal{F}(\Fq^n)_{k_1,<\alpha}|\leq \binom{n}{k_1}_q \frac{q^{k_1}-1}{(q-1)(q^n-1)} (\boldsymbol{b}(\alpha-1)-1).
    \end{equation}
Therefore we have
\[|\mathcal{E}(\Fq^n)_{k_1,\geq \alpha}|=\binom{n}{k_1}_q-|\mathcal{F}(\Fq^n)_{k_1,<\alpha}|\]
and the statement follows using Equation \eqref{eq:Gbis}.
\end{proof}

\begin{proposition}\label{p:countcss}
Let $C_1\subseteq\F_q^n$ be a linear code of parameters $[n,k_1,d_1]$ with $k_1\leq n$, and let $k_2\leq k_1$ and $\beta$ be integers. We have
\[|\mathcal{G}(C_1)_{k_2,\geq \beta}|\geq\binom{k_1}{k_1-k_2}_q\left[1-\frac{q^{n-k_2}-q^{n-k_1}}{(q-1)(q^n-q^{n-k_1})}(\boldsymbol{b}(\beta-1)-1)\right].\]
\end{proposition}
\begin{proof}
Let us denote by $C_1^\perp$ the dual code of $C_1$, which is of dimension $n-k_1$. Note that $\mathcal{G}(C_1)_{k_2,\geq \beta}$ can be rewritten as $\{C_2^\perp\subseteq \Fq^n\mid C_1^\perp\subseteq C_2^\perp, \, \dim(C_2^\perp)=n-k_2, d(C_2^\perp)\geq \beta\}$. Now it is easy to see that  $\mathcal{G}(C_1)_{k_2,\geq \beta}$ is the complement of \[\mathcal{F}(C_1^\perp)_{n-k_2,<\beta}=\{C_2^\perp\subseteq \Fq^n\mid C_1^\perp\subseteq C_2^\perp, \, \dim(C_2^\perp)=n-k_2, \, d(C_2^\perp)< \beta\}\]
in the set \[\mathcal{B}\coloneqq\{C_2^\perp \subseteq \F_q^n \mid C_1^\perp \subseteq C_2^\perp, \, \dim(C_2^\perp)=n-k_2 \},\] which has cardinality $\binom{k_1}{k_1-k_2}_q$.
Applying Proposition \ref{th:count} to the set $\mathcal{F}(C_1^\perp)_{n-k_2,<\beta}$ gives
\begin{equation}\label{eq:G}
    |\mathcal{F}(C_1^\perp)_{n-k_2,<\beta}|\leq \binom{k_1}{k_1-k_2}_q \frac{q^{n-k_2}-q^{n-k_1}}{(q-1)(q^n-q^{n-k_1})} (\boldsymbol{b}(\beta-1)-1).
    \end{equation}
Finally,
\[|\mathcal{G}(C_1)_{k_2,\geq \beta}|=|\mathcal{B}|-|\mathcal{F}(C_1^\perp)_{n-k_2,<\beta}|\]
and the statement follows using Equation \eqref{eq:G}.
\end{proof}
Combining the two previous results, we finally obtain the bound we wanted.

\begin{theorem}\label{th:CSSbound}
Let $\mathcal{S}_{k_1,k_2}^{\alpha,\beta}$ and $\mathcal{S}_{k_1,k_2}$ be the set defined in Equations \eqref{setS'} and \eqref{setS}, respectively. Then
   \[ \frac{|\mathcal{S}_{k_1,k_2}^{\alpha,\beta}|}{|\mathcal{S}_{k_1,k_2}|}\geq \left[1-\frac{q^{n-k_2}-q^{n-k_1}}{(q-1)(q^n-q^{n-k_1})}(\boldsymbol{b}(\beta-1)-1)\right]\left[1-\frac{q^{k_1}-1}{(q-1)(q^n-1)}(\boldsymbol{b}(\alpha-1)-1)\right].\]
\end{theorem}
\begin{proof}
Since  
\[|\mathcal{S}_{k_1,k_2}|=\binom{n}{k_1}_q\binom{k_1}{k_2}_q,\]
the result follows by combining Lemma \ref{countingC1} and Proposition \ref{p:countcss}, using that 
    \[
|\mathcal{S}_{k_1,k_2}^{\alpha,\beta}|=|\mathcal{E}(\Fq^n)_{k_1,\geq \alpha}|\cdot|\mathcal{G}(C_1)_{k_2,\geq \beta}|,
\]
and that $\binom{k_1}{k_1-k_2}_q=\binom{k_1}{k_2}_q$.
\end{proof}

As a corollary of the previous theorem,  we obtain an asymptotic estimate on CSS codes with a fixed minimum distance. In other words, the density of the set $\mathcal{S}_{k_1,k_2}^{\alpha,\beta}$ inside $\mathcal{S}_{k_1,k_2}$.

\begin{corollary}\label{cor:density}
Let $n,k_1,k_2,\alpha,\beta$ be integers such that $k_2\leq k_1\leq n$. Suppose $k_2\geq \beta-1$. Then the set of pairs of linear codes $C_2\subseteq C_1\subseteq \Fq^n$ of dimensions $k_2$ and $k_1$ respectively such that $Q(C_1,C_2)$ is a $\lfloor \frac{d-1}{2}\rfloor$-error-correcting quantum code with $d\geq\min(\alpha,\beta)$ is dense for $q$ large.
\end{corollary}

\begin{proof}
The quantity $\frac{q^{k_1}-1}{(q-1)(q^n-1)}(\boldsymbol{b}(\alpha-1)-1)$ is a polynomial in $q$ with degree $k_1+\alpha-n-2$, which is negative by the Singleton bound. Therefore as $q$ grows we have
\[1-\frac{q^{k_1}-1}{(q-1)(q^n-1)}(\boldsymbol{b}(\alpha-1)-1)\rightarrow 1.\]
The quantity $\frac{q^{n-k_2}-q^{n-k_1}}{(q-1)(q^n-q^{n-k_1})}(\boldsymbol{b}(\beta-1)-1)$ is a polynomial in $q$ of degree
$-k_2+\beta-2.$ Consequently, whenever $k_2\geq \beta-1$ we have \[1-\frac{q^{n-k_2}-q^{n-k_1}}{(q-1)(q^n-q^{n-k_1})}(\boldsymbol{b}(\beta-1)-1)\rightarrow 1 \text{ as } q\rightarrow\infty.\]
The statement follows.
\end{proof}

\begin{remark}
We point out that the condition $k_2\geq \beta-1$ in Corollary~\ref{cor:density} is necessary to have $d_2^\perp\geq \beta$. Indeed, we have
\[0\leq d_2^\perp- \beta\leq -k_2^\perp+n+1-\beta=k_2+1-\beta.\]
\end{remark}

\section{CSS-T Codes: Characterisations and Bounds}\label{s:CSST}\label{sec3}

In the previous sections, we have seen how errors affecting qudits can be represented using the group of tensors of $q\times q$ Pauli matrices, and that CSS codes allow to correct those errors. However, the group of Pauli matrices (or the Clifford group, its normalizer) is not a set of universal quantum gates, that is, it is not enough to approximate any operation that can be performed on a quantum computer. Indeed, universal quantum computation requires the implementation of a so-called non-Clifford gate, and a standard choice for this gate, is the transversal $T$ (Toffoli) gate \cite{T-gate}. Furthermore, it can be shown that a quantum
computer using only operations from the Clifford group can be efficiently simulated on a
classical computer~\cite{S99_nature}. Therefore, the implementation of a non-Clifford gate is a crucial aspect of quantum computing.  In general,  Clifford gates are easier to implement than $T$ gates, and while CSS codes admit a transversal 
implementation of the Clifford group, they don't always support the $T$ gate. To overcome this problem,  Rengaswamy, Calderbank, Newman, and Pfister recently introduced in \cite{RCNP20} a new class of quantum codes derived from the CSS construction, the CSS-T codes, that support the aforementioned transversal $T$ gate as an operator. CSS-T codes are again quantum codes constructed from a pair of classical linear codes that must satisfy some hypotheses (see Definition \ref{def:csst}). Note that, although other quantum codes may support the $T$ gate, it has been proved that among non-degenerate stabilizer codes supporting the $T$ gate transversally, the CSS codes are the optimal ones \cite{RCNP20b}. This is why CSS-T codes are defined as a subfamily of CSS codes.

Few constructions of CSS-T codes are known so far (see \cite{RMCSST} for CSS-T codes from Reed--Muller codes and \cite{CMLMRS24} for CSS-T codes from cyclic and extended cyclic codes), and an open problem in this context, as stated in \cite{RCNP20}, is to find families of CSS-T codes with rate and relative distance both non-vanishing. 
\smallskip

In this section, we focus our attention on CSS-T codes. First, we will generalise their original \emph{binary} definition to any finite field, and give a characterisation of such codes (Theorem~\ref{prop:CSST}). Secondly, we will show that under very natural assumptions, the rate and relative distance of CSS-T codes cannot be simultaneously large (Theorem \ref{th:csstbounds}).

\subsection{$q$-ary CSS-T Codes}
We recall that $\sigma(x)$ denotes the support of the vector $x$ and $\wt(x)=|\sigma(x)|$ its Hamming weight. We also set $[n]\coloneqq \{1,\dots,n\}$. 
The following definition of binary CSS-T codes was introduced in \cite[p.1]{RCNP20}.
\begin{definition}[\cite{RCNP20}]\label{def:csst}
A CSS code $Q(C_1,C_2)$ is called a \emph{(binary) CSS-T code} if the followings hold:
\begin{enumerate}
    \item $C_2$ is an even code; \ie$\wt(x) \equiv 0 \mod 2$ for all $x\in C_2$;
\item for each $x\in C_2$, there exists a self-dual code in $C_1^\perp$ of dimension $\wt(x)/2$ that is supported on $x$, \ie there exists $C_x \subseteq C_1^\perp$ such that $|C_x| = 2^{\wt(x)/2}$, $C_x = C_x^{\perp_x}$, where $\perp_x$ denotes the dual taken in $\Fq^{\wt(x)}$, and for any $z \in C_x$ we have $\sigma(z) \subseteq \sigma(x)$.
\end{enumerate}
\end{definition}

As for the case of CSS codes, the definition of CSS-T code can be extended to non-binary linear codes. The notion of even code is sometimes extended to every finite field in the following way: $C\subseteq \F_{q}^n$ is a \emph{even-like} code if for any $c\in C$ we have
$\sum_{i=1}^n c_i \equiv 0 \mod q$. 
However, in the present paper, we adopt the following definition.
\begin{definition}\label{def:evencodes}
    A linear code $C\subseteq\Fq^n$ is \emph{even} if all of its codewords have even weight.
\end{definition}

\begin{remark}
For $S\subseteq [n]$ let
\[\Fq^n(S)\coloneqq\{y\in\Fq^n\mid \sigma(y)\subseteq S\},\]
and set $\pi_S:\Fq^n\rightarrow\Fq^{|S|}$ to be the projection on the coordinates indexed by $S$.
For the reader familiar with the \emph{puncturing} and \emph{shortening} constructions,
we point out that 
$\pi_S(C)=\punct(C,[n]-S)$, and $\pi_S(C\cap \Fq^n(S))=\short(C,[n]-S).$
The second condition in Definition~\ref{def:csst} can be rephrased as follows:
\begin{enumerate}[(2)]
    \item for any $x\in C_2$, $\pi_{\sigma(x)}(C_1^\perp\cap \mathbb{F}_2^n(\sigma(x)))$  contains a self-dual code.
\end{enumerate}
\end{remark}

We are now ready to give the definition of $q$-ary CSS-T codes.
\begin{definition}\label{def:CSST}
A CSS code $Q(C_1,C_2)$ from two linear codes $C_2\subseteq C_1\subseteq\Fq^n$ is a \emph{$q$-ary CSS-T code} if:
\begin{enumerate}
    \item $C_2$ is an even code;,
\item for any $x\in C_2$ we have that $\pi_{\sigma(x)}(C_1^\perp\cap \Fq^n(\sigma(x)))$ contains a self-dual code.
\end{enumerate}
\end{definition}

In contrast to the binary case, the first condition in the definition of $q$-ary CSS-T codes is not linear. 
Nevertheless, it is necessary to be able to fulfill the second one. 
We shall now focus on the second condition. In \cite{RMCSST} the authors proved that a \emph{binary} code~$C$ contains a self-dual code if and only if its length is even and $C^\perp$ is self-orthogonal. We recall here their nice proof and generalise it to linear codes over $\Fq$ with $q$ a power of $2$. Unfortunately, the proof cannot be generalised further to finite fields of odd characteristic. However, using different techniques, we are able to establish a similar characterisation over such fields; see Proposition~\ref{prop:self-dual} below.

\begin{proposition}\label{prop:felice}
Let $q$ be a power of $2$. A linear code $C\subseteq \F_q^n$ contains a self-dual code if and only if $n$ is even and $C^\perp$ is self-orthogonal, that is, $C^\perp\subseteq C$. 
\end{proposition}
\begin{proof} The requirement on $n$ is clearly necessary for the existence of a self-dual code. 
Suppose that $C$ contains a self-dual code $D$. By dualizing $D\subseteq C$ 
we get $C^\perp\subseteq D^\perp=D \subseteq C$. Therefore~$C^\perp$ is self-orthogonal.
Now let $C^\perp$ be self-orthogonal and let $n$ be even. Consider the set
\[\mathcal{L}\coloneqq\{D\subseteq C \mid C^\perp \subseteq D \text{ and }\dim D=n/2\}.\] Let $k$ denote the dimension of $C$. By \cite[Proposition~2.5]{BR20} we have \[|\mathcal{L}|=\binom{k-(n-k)}{\frac{n}{2}-(n-k)}_q=\binom{2k-n}{k-\frac{n}{2}}_q=\prod_{i=0}^{k-\frac{n}{2}-1} \frac{q^{2k-n-i}-1}{q^{i+1}-1}.\] Since $q$ is even, $|\mathcal{L}|$ is odd. Observe that if $D\in\mathcal{L}$, then $D^\perp\in\mathcal{L}$. Thus $\mathcal{L}$ must contain at least one code $D$ such that $D=D^\perp$, which concludes the proof.
\end{proof}

We now turn to the case where $q$ is odd.

\begin{proposition}\label{prop:self-dual}
Let $C\subseteq\Fq^n$ be a linear code of dimension $k\geq n/2$. 
\begin{enumerate}
\item If $q\equiv 1\mod 4$, then $C$ contains a self-dual code if and only if $n$ is even and $C^\perp$ is self-orthogonal.
\item If $q\equiv 3\mod 4$, then $C$ contains a self-dual code if and only if $n \equiv 0 \mod 4$ and~$C^\perp$ is self-orthogonal.
\end{enumerate}
\end{proposition}

\begin{proof}
If $C \subseteq \F_q^n$ contains a self-dual code, say $A$, then $\dim(A)=n-\dim(A)$ implies that $n$ is even. Moreover, 
$C \supseteq A$ implies
$C^\perp \subseteq A^\perp=A \subseteq C=(C^\perp)^\perp$, hence $C^\perp$ is self-orthogonal. We study the two cases separately, using results on finite bilinear spaces revisited and extended in~\cite{AlbertoHeide}.

    \begin{enumerate}
        \item Suppose $q\equiv 1 \mod 4$. If $C$ contains a self-dual code then we have already shown that $n$ is even and $C^\perp$ is self-orthogonal. Now suppose that $n$ is even and that $C^\perp$ is self-orthogonal. We want to prove that there exists~$A$ self-orthogonal with $\dim(A)=n/2$ and 
        $\F_q^n \supseteq A \supseteq C^\perp$. If $k=n/2$ then $C$ is self-dual, and the statement follows. So, let us consider the case $k>n/2$. By~\cite[Proposition 2.9]{AlbertoHeide}, such a space exists if and only if there exists a subspace $A' \subseteq C/C^\perp$ of dimension $n/2-(n-k)=k-n/2$
        that is self-orthogonal with respect to the non-degenerate symmetric bilinear form induced by the standard inner product on the quotient $C/C^\perp$. Since $q \equiv 1 \mod 4$, $(\F_q^n, \cdot)$ has type (H); see~\cite[Proposition~3.2]{AlbertoHeide}.  Since $k>n/2$ by assumption, we have $n-k \neq n/2$ and therefore by the second part of~\cite[Proposition~2.9]{AlbertoHeide}, the induced bilinear space~$C/C^\perp$ also has type (H). In particular, by~\cite[Definition 2.8]{AlbertoHeide} and the definition of Witt index, it contains a self-orthogonal subspace of dimension $\dim(C/C^\perp)/2=k-n/2$, as desired.

        \item Suppose $q\equiv 3 \mod 4$. If $C$ contains a self-dual code then we have already shown that $n$ is even and $C^\perp$ is self-orthogonal. We need to show that $n \equiv 0 \mod 4$ as well. If $k=n/2$ then this follows from \cite[\S 2.4.3]{NRS06}. Hence, let us suppose $k>n/2$.
        As in the previous part of the proof $C$ containing a self-dual code implies the existence of a self-dual code in the quotient bilinear space $C/C^\perp$.
        Since $q\equiv 3 \mod 4$, $n$ is even, and $k>n/2$, 
        by~\cite[Propositions~2.9 and~3.2]{AlbertoHeide}, the latter space has type (H) if $n\equiv 0 \mod 4$ and type (E) if 
        $n\equiv 2 \mod 4$. By~\cite[Definition 2.8]{AlbertoHeide} and the definition of Witt index, the existence of a self-dual code is only possible when the type is (H), which then gives $n\equiv 0 \mod 4$. The same argument shows that if $n \equiv 0 \mod 4$ and~$C^\perp$ is self-orthogonal, then~$C$ contains a self-dual code.    \qedhere
    \end{enumerate}
\end{proof}

Thanks to Proposition \ref{prop:self-dual} we can provide the following characterisation of CSS-T codes.

\begin{theorem}\label{prop:CSST}
Let $C_2\subseteq C_1\subseteq\Fq^n$ be linear codes and let $C_2$ be even. 
\begin{enumerate}
    \item If $q$ is even or $q\equiv 1 \mod 4$, then $Q(C_1,C_2)$ is a CSS-T code if and only if for any $x\in C_2$ we have 
\[\pi_{\sigma(x)}(C_1)\subseteq\pi_{\sigma(x)}(C_1)^{\perp_x}.\]
\item If $q\equiv 3\mod 4$, then $Q(C_1,C_2)$ is a CSS-T code if and only if for any $x\in C_2$ we have 
\[\pi_{\sigma(x)}(C_1)\subseteq\pi_{\sigma(x)}(C_1)^{\perp_x} \quad
\text{and }\quad \frac{\wt(x)}{2}\equiv 0 \mod 2.\]
\end{enumerate}

\end{theorem}
\begin{proof} Let us fix $x\in C_2$. We know from Propositions \ref{prop:felice} and \ref{prop:self-dual} that when $q$ is even or $q\equiv 1 \mod 4$ then $\pi_{\sigma(x)}(C_1^\perp\cap \Fq^n(\sigma(x)))$ contains a self-dual code if and only if its length is even and $(\pi_{\sigma(x)}(C_1^\perp\cap \Fq^n(\sigma(x))))^{\perp_x}$ is self-orthogonal, and that when $q\equiv 3 \mod 4$ we furthermore need the condition $\frac{\wt(x)}{2}\equiv 0 \mod 2$. Since $C_2$ is an even code we have that $\wt(x)$ is even. Observe that $(\pi_{\sigma(x)}(C_1^\perp\cap \Fq^n(\sigma(x))))^{\perp_x}=\pi_{\sigma(x)}((C_1^\perp\cap \Fq^n(\sigma(x)))^\perp)$. 
From \cite[Theorem 1.5.7 (i)]{HP10} we know that 
\[(\pi_{\sigma(x)}(C_1^\perp\cap \Fq^n(\sigma(x))))^{\perp_x}=\pi_{\sigma(x)}(C_1),\]
which concludes the proof.
\end{proof}

\subsection{Rate and Relative Distances of CSS-T Codes}

As we have already mentioned, in~\cite{RCNP20} the authors pose the open problem about the existence of 
families of CSS-T codes with non-vanishing asymptotic rate and asymptotic relative distance.  
In this subsection, we focus our attention on these parameters.
We start by recalling a well-known property of punctured codes (cf. \cite[Proposition~2.1.4]{PWBJ18}).

\begin{proposition}\label{prop:punct}
    Let $C$ be a $[n,k,d]$ code. Let $I$ be a subset of $\{1,\dots,n\}$ of size $r$. Then, if $n-r< d$, $\pi_I(C)$ is a $[r,k,\geq d-(n-r)]$ code.
\end{proposition}

In the sequel we follow the notation of the 
previous subsections and also set
\[R=\frac{k_1-k_2}{n},\quad
    \delta_1=\frac{d_1}{n},\quad
   \delta_2^\perp=\frac{d_2^\perp}{n},\quad
    \delta=\min(\delta_1,\delta_2^\perp).\]
Since $d_2^\perp \geq \min(d_1,d_2^\perp)$, we have $\delta\leq \delta_2^\perp$.

\begin{theorem}\label{th:csstbounds}
    Let $C_2\subseteq C_1\subseteq\Fq ^n$ be such that $Q(C_1,C_2)$ is a CSS-T code. Then the following facts hold.
    
    \begin{enumerate}
        \item\label{bound:uno} If there exists a codeword $x\in C_2$ such that $\wt(x)\geq n -  k_2+1$, then
    \[R+\frac{\delta_2^\perp}{2} \leq \frac{1}{2}.\]
    In particular, $R+\delta/2 \leq 1/2$.
     \item\label{bound:due} If there exists a codeword $x\in C_2$ such that $\wt(x)>n -  d_1$, then 
    \[R+\delta_2^\perp \leq \frac{1}{2}+\frac{1}{n}.\]
    In particular, $R+\delta \leq 1/2+1/n$.
     \item\label{bound:duebis} If there exists a codeword $x\in C_2$ such that $\wt(x)=n -  d_1$, then 
    \[R+\frac{\delta_1}{2}+\delta_2^\perp \leq \frac{1}{2}+\frac{2}{n}.\]
    In particular, $R+3\delta/2 \leq 1/2+2/n$.
    \end{enumerate}
\end{theorem}
\begin{proof}
\begin{enumerate}
\item By Proposition \ref{prop:self-dual} since $Q(C_1,C_2)$  is a CSS-T code, for any $x\in C_2$ the code $\pi_{\sigma(x)}(C_1)$, which has length $\wt(x)$, is self-orthogonal. Therefore, we must have $k_x\coloneqq\dim \pi_{\sigma(x)}(C_1) \leq \wt(x)/2\leq n/2$. 
 Clearly we have $k_x=k_1-\dim\left(\ker\pi_{\sigma(x)}\cap C_1\right)\geq k_1-\dim\left(\ker\pi_{\sigma(x)}\right)= k_1-(n-\wt(x))$, which rewrites as
 \[k_1\leq k_x+n-\wt(x)\leq n-\frac{\wt(x)}{2}.\]
 By our assumptions, there exists at least one codeword $x\in C_2$ such that $\wt(x)\geq n-k_2+1$. This implies
 $k_1\leq (n+k_2-1)/2$
 and thus we have
 \[k_1-k_2\leq \frac{n-k_2-1}{2}.\]
 By the Singleton bound we also have
     $d_2^\perp-1\leq n-k_2^\perp=k_2.$
     These bounds together give
     \[k_1-k_2\leq \frac{n-d_2^\perp}{2}. \]
We conclude by dividing both terms by $n$.

\item By assumption, we have $\wt(x)>n-d_1$, therefore there exist a positive integer $\lambda$ such that $\wt(x)=n-d_1+\lambda$.  Since $n-\wt(x)< d_1$, by Proposition \ref{prop:punct}, $\pi_{\sigma(x)}(C_1)$ has dimension $k_1$. By Proposition \ref{prop:self-dual} since $Q(C_1,C_2)$  is a CSS-T code, the code $\pi_{\sigma(x)}(C_1)$, which has length $\wt(x)$, is self-orthogonal. Therefore, we must have $k_1\leq \wt(x)/2$. Furthermore, we have $d_2^\perp-1\leq n-k_2^\perp=k_2.$
     In sum, this gives
     \[k_1-k_2\leq \frac{n -  d_1+\lambda}{2}-d_2^\perp+1.\]
Since $\lambda$ is necessarily strictly smaller than $d_1$, we have
 \[k_1-k_2\leq \frac{n}{2}-d_2^\perp+1.\]
Dividing by $n$ we get the result.
\item Since $\wt(x)= n -  d_1$, we have that $\pi_{\sigma(x)}(C_1)$ has dimension $k_1-1$ or $k_1$. In the first case, using Proposition \ref{prop:self-dual} we know that
\[k_1-1\leq \frac{\wt(x)}{2}-1=\frac{n-d_1}{2}-1.\]
Moreover by the Singleton bound we also have
     $d_2^\perp-1\leq n-k_2^\perp=k_2.$
     Thus, we obtain
     \[k_1-k_2\leq \frac{n-d_1}{2}-d_2^\perp+2. \]
Dividing by $n$ we get
\[R+\frac{\delta_1}{2}+\delta_2^\perp \leq \frac{1}{2}+\frac{2}{n}.\]

In the second case, we can perform exactly the same reasoning, using now that $k_1\leq \wt(x)/2= (n-d_1)/2$. Therefore we conclude 
\[R+\frac{\delta_1}{2}+\delta_2^\perp \leq \frac{1}{2}+\frac{1}{n}\leq \frac{1}{2}+\frac{2}{n},\]
\end{enumerate}
as desired.
\end{proof}

\begin{remark}
A recent paper \cite{RMCSST} proposes a construction of CSS-T codes from binary Reed-Muller codes with asymptotic rate $1/2$ and vanishing asymptotic relative distance.
    If we consider Reed-Muller codes $C_1=\RM(r',m)$ and $C_2=\RM(r,m)$ with $r'>r$, then since the codeword $\boldsymbol{1}\in C_2$, the bound \eqref{bound:uno} of Theorem \ref{th:csstbounds} applies to $Q(C_1,C_2)$. In particular, this means that binary CSS-T codes from binary Reed-Muller codes cannot have rate bigger than $1/2$, and when they have rate exactly equal to $1/2$ they must have zero minimum distance.  Therefore our bound shows that the construction in \cite{RMCSST} is optimal.
\end{remark}

\begin{remark}  Consider the code $Q(C^\perp,C)$ where $C$ is a (strictly) self-orthogonal code,  that is $C\subsetneq C^\perp$. If $Q(C^\perp,C)$ is a CSS-T code, then the assumption $\wt(x)> n-d_1$ is never satisfied. Indeed, we have seen that if $Q(C_1,C_2)$ is a CSS-T code, then the existence of a codeword $x\in C_2$ such that $\wt(x)> n-d_1$ implies $k_1\leq \frac{n}{2}$. However, for $C$ to be (strictly) self-orthogonal we must have $k_1>\frac{n}{2}$.
\end{remark}

The results of Theorem~\ref{th:csstbounds} show that the parameters $R$ and $\delta$ of a CSS-T code $Q(C_1,C_2)$ must satisfy certain bounds if the code $C_2$ contains a codeword of a certain weight. 
In the following proposition, we show that the existence condition in point \eqref{bound:uno} of the theorem is likely satisfied for a linear code over $\Fq$ for large $q$.

\begin{proposition}
    Let $k,n,\omega$ be positive integers such that $k<n$ and $\omega >n-k+1$. For any prime power $q$, consider the set 
    \[\mathcal{L}_q=\{ C\subseteq\Fq^n \mid \dim C =k, \, \exists x\in C \mbox{ with } \wt(x)=\omega\}.\]
    Then, for $q\rightarrow\infty$ we have
    \[\frac{|\mathcal{L}_q|}{\binom{n}{k}_q}\rightarrow 1.\]
    \end{proposition}

\begin{proof}
The set $\mathcal{L}_q$  is the set of $q$-ary linear codes of length $n$ and dimension $k$ that contain a codeword of weight $\omega$.
Since  $\smash{\binom{n}{k}_q}$ is the number of $[n,k]_q$ linear codes, 
we have 
$\smash{|\mathcal{L}_q|\leq 
\binom{n}{k}_q\sim q^{(n-k)k}}$. We show that, asymptotically, also $\smash{|\mathcal{L}_q|}$ grows as $q^{(n-k)k}$.

We consider a bipartite graph with on the one side the set $K$ of vectors $v\in\Fq^n$ that have weight exactly $\omega$, and on the other side the set of $[n,k]_q$ linear codes. 
We connect each code to the vectors it contains.
Let us set $\beta=\frac{|K|-1}{q-1}=\binom{n}{\omega}(q-1)^{\omega-1}$. In this setting, we can apply~\cite[Theorem 3.8]{GR22} and obtain 
    \[|\mathcal{L}_q| \geq \frac{\beta \binom{n-1}{k-1}_q}{1+(\beta-1)\frac{q^{k-1}-1}{q^{n-1}-1}}.\]

Asymptotically, for $q\rightarrow\infty$, the numerator grows as  $\smash{\binom{n}{\omega}q^{\omega-1+(k-1)(n-k)}}$.
   Regarding the denominator,
   since $\smash{(\beta-1)\frac{q^{k-1}-1}{q^{n-1}-1}\sim \binom{n}{\omega}q^{\omega-1+k-n}}$ as $q\rightarrow\infty$, the right summand of the denominator is bigger than $1$ as soon as  $\omega-1+k-n >0$. Therefore, under our assumption on~$\omega$, for $q\rightarrow\infty$ we have
    \[|\mathcal{L}_q| \geq q^{(n-k)k}\sim \binom{n}{k}_q,\]
which concludes the proof.
\end{proof}

\begin{remark}\label{remark-CSSTrate}
   We can construct a CSS-T code attaining the first bound of Theorem~\ref{th:csstbounds} whenever there exists a self-dual code with a codeword of full support of even weight, in the following way. Let $C_1\subseteq\Fq^n$ be a self-dual code, and let $x\in C_1$ be an even weight codeword of full support, i.e., $x_i\neq0$ for all $i=1,\dots,n$. Take $C_2=\langle x \rangle$. Then we have $C_2\subset C_1$, $C_2$ is even, and $d_2^\perp=2$. The first two statements are trivial, and the third one easily follows by observing that by the Singleton bound one has $d_2^\perp \leq 2$, and that since $x$ has full support $C_2^\perp$ does not contain words of weight $1$.
   Moreover, since $x$ has full support and $C_1=C_1^\perp$, we have that $\pi_\sigma(x)(C_1^\perp\cap \Fq^n(\sigma(x))=C_1^\perp=C_1$ contains a self-dual code. Therefore, by Definition \ref{def:CSST}, $Q(C_1,C_2)$ is a CSS-T code. Its rate satisfies
   \[R=\frac{k_1-1}{n}=\frac{n-2}{2n}=\frac{1}{2}-\frac{1}{n}.\]
   Therefore, we have
   \[R+\frac{\delta_2^\perp}{2}=\frac{1}{2}-\frac{1}{n}+\frac{1}{n}=\frac{1}{2}.\]
\end{remark}

\section{A construction of CSS-T codes from the Hermitian curve}\label{s:AGCSST}\label{sec4}

In this last section, we show a simple construction of CSS-T codes using the Hermitian curve. This construction is optimal with respect to the asymptotic bound \eqref{bound:uno} of Theorem \ref{th:csstbounds}.
We will use standard notations and results on AG codes. We refer the reader to \cite{Stepanov, Stich09} for more details and proofs.

From now on, we let $q$ be a power of $2$ and we consider  the Hermitian curve $\mathcal{C}$ defined over~$\mathbb{F}_{q^2}$ by the equation \[
    y^q+y=x^{q+1}.
    \]
 By Pl\"ucker's formula, $\mathcal{C}$ has genus $g=\frac{1}{2}q(q-1)$. It has a unique point at infinity, which we denote by $P_\infty$, and $n=q^3$ affine points over $\mathbb{F}_{q^2}$. In particular, the $q^3$ affine points are characterized as follows \cite[Lemma 6.4.4]{Stich09}: for any $\alpha\in \F_{q^2}$, there exist $q$ elements $\beta\in\F_{q^2}$ such that $P_{\alpha,\beta}=(\alpha,\beta)$ is an affine point of $\mathcal{C}$.
 
 We will consider one-point codes on $\mathcal{C}$ of the form 
 \[
 C(mP_\infty,D)\coloneqq\{ev_D(f)=(f(P_1),\dots,f(P_n))\in\Fq^n\mid f\in L(mP_\infty)\},
 \]
where $m>0$ is an integer, $D=\sum_{P\in \mathcal{C}(\mathbb{F}_{q^2})\setminus\{P_\infty\}} P$ is the sum of all affine points of $\mathcal{C}$, and $L(mP_\infty)$ is the Riemann--Roch space of rational functions $f$ on $\mathcal{C}$ such that $\mathrm{div}f+mP_{\infty}\geq0$.

\begin{lemma}\label{lemma:evensubcode}
    Let $C_1$ be the code $C(mP_\infty,D)$ on $\mathcal{C}$. The subcode $C_2\subseteq C_1$ constructed by evaluating the functions in $L\coloneqq\langle x^i : 0\leq i \leq m/q\rangle$ at the points of $D$ is even. Furthermore, it is the largest subcode of $C_1$ that is even.
\end{lemma}
\begin{proof}
 Without loss of generality, we can reorder the $q^3$ rational points in the support of $D$ in $q^2$ sets of cardinality $q$, each made of points with the same first coordinate. Take $f\in L$. Then, the coordinates of the codeword $ev_D(f)$ can be divided into $q^2$ sets of cardinality $q$, the $q$ coordinates in each set being the same. In particular, one coordinate in a set is zero if and only if the other $q-1$ coordinates in the same set are zero. Since $q$ is even, this shows that the weight of  $ev_D(f)$ is even. Now, let us prove that $L$ is the maximal subset of $L(m P_\infty)$ giving an even code. First, let us remark that the bound $i\leq m/q$ cannot be enlarged \cite[Lemma 6.4.4(e)]{Stich09}. Secondly, there is only one point on $\mathcal{C}$ with $y=0$, thus the function $y$ and its powers lead to codewords of odd weight. Finally, we consider functions of the form $x^iy^j$, for $i,j\in\N$. Without loss of generality, let us consider $xy$. Fix $x=1$, then there are at least two elements $y_1$ and $y_2$ such that $(1,y_1)$ and $(1,y_2)$ are points on $\mathcal{C}$. Let $z$ be the generator of the extension $\mathbb{F}_{q^2}/\Fq$. Then there exists a choice of integers $j_i$ such that $\smash{y_1=\sum_{i=0}^{m/q} z^{j_i}}$, that is $(1,y_1)$ is a zero of the function $\smash{f=\sum_{i=0}^{m/q} z^{j_i}x^i+xy\in \langle xy,x^i : 0\leq i \leq m/q\rangle}$, and any $(1,y)$ with $y\neq y_1$ is not a zero of $f$. In a similar way, if $(\alpha,\beta_1)$ is a root of $f$, then there is no other~$\beta_i$ such that $(\alpha,\beta_i)$ is a root of $f$. Therefore, the weight of $ev_D(f)$ is odd. This concludes the proof.
\end{proof}
\begin{lemma}\label{lem:dualmindist}
    Let $C_2$ be the linear code from Lemma~\ref{lemma:evensubcode}. Then its dual code $C_2^\perp$ has minimum distance $d_2^\perp=2$.
\end{lemma}
\begin{proof}
    From the reasoning in the proof of Lemma \ref{lemma:evensubcode}, we see that every codeword $c\in C_2$ has~$q^2$ coordinates repeated $q$ times. More specifically, let $i\neq j$ be two positions where $c_i=c_j$. Then, for any other codeword $c'\in C_2$ we have $c'_i=c'_j$. Therefore, since we are working in characteristic $2$, the codeword $e_i+e_j$, where $e_i$ denotes the $i$-th vector of the standard basis, belongs to $C_2^\perp$, and thus we have $d_2^\perp \leq 2$. Since $\boldsymbol{1}\in C_2$ we also have $d_2^\perp >1$, which concludes the proof.
\end{proof}

    By Lemma \ref{lem:dualmindist}, the maximal dimension even subcode $C_2$ of $C(m P_\infty,D)$ has the same dual minimum distance of the code generated by the one vector $\boldsymbol{1}$, and of course bigger dimension. This justifies why in the following, we will consider $C_2=\langle \boldsymbol{1}\rangle$ for constructing Hermitian CSS-T codes.

\begin{theorem}\label{th:hermitianCSST}
Let $q$ be a power of $2$ and consider the Hermitian curve $\mathcal{C}$ of genus $g=\frac12q(q-1)$ defined over $\mathbb{F}_{q^2}$.
    Set $q^2-q-1\leq m\leq (q^3+q^2-q-2)/2$ and let $D=\sum_{P\in \mathcal{C}(\mathbb{F}_{q^2})\setminus\{P_\infty\}} P$. Let $C_1=C(mP_\infty, D)$ and let $C_2=\langle \boldsymbol{1}\rangle $. Then, the quantum code $Q(C_1,C_2)$ is a $\llbracket q^3,m-g\rrbracket_q$ CSS-T code. In particular, for $m=(q^3+q^2-q-2)/2$, the rate and relative distance of $Q(C_1,C_2)$ attain bound \eqref{bound:uno} of Theorem \ref{th:csstbounds}.
\end{theorem}
\begin{proof}
The dimension $k_1$ of $C_1$ is easily computed via the Riemann--Roch theorem and is $k_1= m - g +1$.
    By construction $C_2\subseteq C_1$, thus $Q(C_1,C_2)$ is a CSS code. Its dimension is $k_1-k_2= m-g$ and the minimum distances are given by $d_1\geq q^3-m$ and $d_2^\perp=2$. By construction, $C_2$ is an even code and for any $x\in C_2$, we have $\pi_{\sigma(x)}(C_1)=C_1$. Since $2m \leq n+2g-2 $, $C_1$ is self-orthogonal \cite[Proposition~8.3.2]{Stich09}. Its length is even by construction. Hence, by Proposition~\ref{prop:CSST}, $Q(C_1,C_2)$ is a CSS-T code. The last statement follows by a simple calculation.
\end{proof}

\begin{example}
    Following Theorem \ref{th:hermitianCSST}, we can obtain explicitly the parameters of CSS-T codes from the Hermitian curve once $q$ is fixed. For instance, for $q=2$ we get a $\llbracket 8,3\rrbracket_2$ CSS-T code, and for $q=4$ we have a $\llbracket 64,31\rrbracket_4$ CSS-T code.
\end{example}

\section*{Conclusions} In this paper,
we investigated CSS and CSS-T codes for quantum error correction. We described the typical behaviour of a CSS and of a CSS-T code with respect to correction capability, showing that, over a large field, a uniformly random CSS code has good distance properties, while a CSS-T code hasn't. We also established bounds on the parameters of a CSS-T code, evaluating the performance of previous constructions. We then provided a simple construction of CSS-T codes from Hermitian curves. The paper also offers a concise introduction to quantum error correction from the point of view of classical coding theory.

\section{Declarations}
\subsection*{Ethical Approval}
This declaration is not applicable.

\subsection*{Funding}
E. Berardini was supported by the 
 EU Horizon 2020 research and innovation programme under the MSCA 
 grant 899987 until April 2023, and by the grant ANR-21-CE39-0009-BARRACUDA of the French National Research Agency. 
A. Caminata is supported by the Italian PRIN2020 grant 2020355B8Y ``Squarefree Gr\"obner degenerations, special varieties and related topics'',  by the Italian PRIN2022 grant 2022J4HRR ``Mathematical Primitives for Post Quantum Digital Signatures'', by the INdAM--GNSAGA grant ``New theoretical perspectives via Gr\"obner bases'', by the MUR Excellence Department Project awarded to Dipartimento di Matematica, Università di Genova, CUP D33C23001110001, and by the European Union within the program NextGenerationEU. 
A. Ravagnani is supported by the Dutch Research Council through grants VI.Vidi.203.045 and OCENW.KLEIN.539, 
by the Royal Academy of Arts and Sciences of the Netherlands, and by the Horizon Europe MSCA-DN grant ``ENCODE''.

\subsection*{Availability of data and materials}
This declaration is not applicable.

\bibliographystyle{siam}
\bibliography{quantum_biblio}

\end{document}